\documentstyle[12pt]{article}
\begin{document}

\setcounter{page}{1}
\newcommand{\newc}{\newcommand}
\newc{\ra}{\rightarrow}
\newc{\lra}{\leftrightarrow}
\newc{\beq}{\begin{equation}}
\newc{\eeq}{\end{equation}}
\newc{\barr}{\begin{eqnarray}}
\newc{\err}{\end{eqnarray}}
\author{E. G. Floratos$^{(*)}$ \\
Institute of Nuclear Physics, \\
NSRC  Demokritos,\\
  Athens, Greece\\
and\\
G. K. Leontaris$^{(**)}$ \\
Centre de Physique Theorique, Ecole Polytechnique,\\
F-91128 Palaiseau, France}

\title{Analytic Approach to the MSSM Mass  \\
Spectrum in the Large Tan$\beta$ Regime.}
\date{}
\maketitle

\begin{abstract}
In various unified extensions of the Minimal Supersymmetric
Standard Model, the Yukawa couplings of the third generation
are predicted to be of the same order.  As a result, low energy
measured mass ratios, require large ratios of the standard model
Higgs vacuum expectation values, corresponding to a large value
of the parameter $tan\beta$. We present  analytic solutions
for the Yukawa couplings and the Higgs and third generation squark
masses, in the case of large top and bottom Yukawa couplings.
We examine regions of these Yukawas which give  predictions for
the top mass compatible with the present experimentally
determined top mass and provide useful approximate  formulae for
the scalars. We discuss the implications on the Radiative
Symmetry Breaking mechanism and derive constraints on the
undetermined initial conditions of the scalars.

\end{abstract}

\thispagestyle{empty}
\vfill
\noindent IOA 320/95

\noindent hep-ph/9503455

\vspace{.5cm}
\hrule
\vspace{.3cm}
{\small
\noindent
$^{(*)}$On leave of absence from Phys. Dept., University of Crete,
Iraklion,Crete, Greece}\\
{\small
$^{(**)}$On leave of absence from Phys. Dept., University of Ioannina,
451 10 Ioannina, Greece}

\newpage

\section{Introduction}

If one adopts the idea of unification  of all forces, then the
most natural candidate which is at the same time the  simplest
extension of the standard model (SM) of strong and electroweak
interactions, is the minimal supersymmetric standard model (MSSM).
Indeed, the low energy measured values
of the three gauge couplings are consistent with simple unification
values at an energy of ${\cal O}(10^{16}GeV)$, provided that the
SM -- spectrum is extended to that of the MSSM at an energy scale
$M_{SUSY}^{eff}\sim 1 TeV$. Furthermore, unified supersymmetric models
solve  successfully the hierarchy  problem\cite{susy}
of their corresponding non--supersymmetric versions.
Recently, in view of the possible observation of supersymmetric
particles in the upcoming colliders, there has been a growing interest
in the determination of the complete mass spectrum of MSSM.
Nevertheless, although the MSSM is considered as a successful extension
of the SM, it still has a number of arbitrary parameters which are not
fixed by the theory. It is expected however that these parameters could
be reduced by embedding the MSSM in a more fundamental theory such
as supergravity or superstrings. In the present status of MSSM and its
 Grand Unified extensions, (SUSY--GUTs),
the Yukawa couplings and the initial conditions for the scalar mass
terms are treated as arbitrary parameters of the theory.
Most of the SUSY--GUTs predict certain relations between the Yukawa
couplings of the third generation, up to an overall constant which
is left arbitrary. Certain string constructions may relate this
constant to the unified value of the common gauge coupling at the
string scale,  usually up to a factor of ${\cal O}(1)$.
Most common relations that hold in these  models\cite{SU(5/4} are
\beq
h_t\sim h_b\sim  h_{\tau},\,\, h_t > h_b \sim h_{\tau}  \,\,
 or \,\, h_t\sim h_b > h_{\tau}
\eeq
Remarkably, it was shown by numerical analyses\cite{top5/4},
that the top mass is predicted  to be  $m_t\sim 180GeV$ which
is compatible with the recent experimental findings\cite{cdf}.
Furthermore, in the context of MSSM, the top mass has been calculated
in the recent literature mainly by numerical  methods or
semi--analytically  in the case of the exact equality  $h_t = h_b$
\cite{cpw,kpz2}. In the above cases, the experimentally observed low
energy  hierarchy $m_t\gg m_b$ demands a large  ratio of the two
Higgs  vacuum expectation values (vevs), $tan\beta
\equiv \frac {\upsilon_2}{\upsilon_1} \gg 1$.

As far as the scalar masses are concerned, the aforementioned theories
have not been  very predictive yet. In the simplest treatments,
one usually assumes that all scalar masses (i.e. higgses and soft
susy masses ) have the same boundary condition at the unification scale
(universality), while their splittings at low energies arise mainly
from the renormalisation group running. In general, however, one expects
that supergravity or superstrings will induce non--universal conditions
for the scalar masses.

In a certain class of supergravity models
\cite{nosc}, each scalar mass $\tilde m_i$ can be parametrised in terms
of a coefficient  $\xi_i\sim {\cal O}(1)$ (calculable in specific models),
and the supersymmetry breaking scale which is proportional to the
gravitino mass $m_{3/2}$  corresponding to an approximately flat
direction of the underlying fundamental theory.
In the effective low energy theory, the over all scale -- set by the
gravitino mass $m_{3/2}$ -- is not known, since  $m_{3/2}$ depends on
singlet scalar field vevs of the higher unknown theory. It has been
claimed however, that since $m_{3/2}$ is a field dependent quantity,
its value can be determined dynamically by minimizing \cite{kpz} the
vacuum energy of the effective  low energy theory with respect to
$m_{3/2}$. This procedure has shown that under general assumptions
for the higher theory and a wide choice of $\xi_i$'s\cite{lt1},
the $m_{3/2}$ scale is of the required order, i.e. $\sim {\cal O}(m_Z)$.
This tells us that supersymmetry is broken at the right scale
 to protect scalars from large radiative corrections,  while
the resulting susy scalar spectrum is accessible by the future
colliders.

While  non--universality  is  not an inescapable
prediction  of the fundamental theory,  it is a necessity
for large $tan\beta$. Indeed, if we are to take seriously
 predictions of exact or approximate $h_t,h_b$ equality at $M_U$,
the radiative breaking of the $SU(2)\times U(1)$ gauge symmetry
\cite{rsb} cannot  be realized naturally if we assume universal
boundary conditions -- in particular for the higgs mass parameters.

In the present work, we wish to explore the region of the parameter
space of  $\xi_i=\tilde m_i^2/m_{3/2}^2,...$ which naturally leads
to radiative symmetry breaking of the electroweak gauge group,
protecting simultaneously the scalar quarks from receiving negative
masses. This latter phenomenon is quite possible due to large negative
radiative corrections on the scalars of the third generation.
We should also point out that another possible consequence of using
non--universal boundary conditions for scalars\cite{nonuni}
is the appearance of enhanced flavor changing neutral currents
(FCNCs)\cite{Luis}. To suppress FCNCs within the experimentally
accepted limits,
it is adequate to consider only cases diagonal in the flavor space.
In our analysis we are going to derive semi--analytic formulae which
will prove  useful as will be explained  in the next section.

Our paper is organized as follows: In section 2, in order for the
paper to be self contained, we start with a short review of the low
energy effective supersymmetric theory and explain our notation.
We  proceed by presenting the analytic solution for large $h_t-h_b$
couplings and comment on their applicability to the minimization
conditions of the low energy potential with respect to higgs vevs
and the gravitino mass. In section 3 we apply our solutions to derive
analytic expressions for various useful quantities (including tan$\beta$).
Section 4 deals with the coupled differential equations for scalar mass of
the third generation in the regime of large tan$\beta$. It is shown
that the solution of the system of equations for all third generation
scalar masses is reduced to one simple first order equation of Riccati
type. We derive analytic solutions in the limiting cases of low and
large $tan\beta$ cases and explore the non--universal conditions which
are compatible with RSB. Finally, in section 5 we present our numerical
results and conclusions.

\newpage

\section{Large Top-Bottom Couplings in MSSM.}

The theory which concerns us here is a supersymmetric low energy
effective theory based on the general structure of
spontaneously broken $N=1$ supergravity \cite{sbs}.
Although such a theory still leaves a lot of arbitrariness,
in the minimal case the effective  theory can be described
in terms  of five new parameters only, namely the gravitino
mass  term  $ m_{3/2}$  the $m_{1/2}$, $\mu$ , $m_3^2 = B \mu$
mass parameters and  the trilinear parameter  $A$.
In the present paper however, for reasons explained already in the
introduction, we will assume non--universality for the higgs and
the other scalar mass  parameters at the unification scale $M_U$,
thus our parameter space  will be enlarged.
At the tree level, the masses of the fermions are
obtained from the trilinear superpotential given by
\begin{equation}
{\cal W}=h_Q^{ij}Q_iu_j^cH_2 + h_D^{ij}Q_id_j^cH_1
+ h_L^{ij}L_ie_j^cH_1+ \mu H_1H_2\label{sup}
\end{equation}
Supersymmetry is broken softly through the
`` soft -- breaking '' terms which provide with masses all the
s--particles. Thus, the gauginos $\lambda_i$ receive  masses from
terms $\frac 12 \tilde m_i\lambda_i\bar \lambda_i$.
Scalar mass terms arise from the effective potential which
in the minimal case is  of the form
\begin{equation}
{\cal L}=\Sigma |\frac{\partial {\cal W}}{\partial\phi_i}|^2 +
\bar A m_{3/2}({\cal W} + {\cal W}^*) +
 m_{3/2}^2 \Sigma {\mid \phi_i\mid}^2+\bar B  m_{3/2}
( \Sigma \phi_i\frac{\partial {\cal W}}{\partial\phi_i}+ h.c.)+...
\label{qupo}
\end{equation}
where the $\{ ... \}$ represent $D$--term contributions.
$\bar A , \bar B$ are  parameters of ${\cal O}(1)$ and in the
case of non--universality they may have different values for each
superpotential term, i.e. $\bar A = \{\bar A_Q , \bar A_D, ...\}$ etc.
Thus, generalizing the Lagrangian in (\ref{qupo})
 for the non--universal case, one obtains soft
squark and slepton mass terms of the form $\tilde m_n^2
\tilde q_n^*\tilde q_n$,
where $ q_n =\{  Q_n , u^c_n , d^c_n , L_n , e^c_n , \nu_n\}$.
Moreover, due to the trilinear  superpotential couplings in (\ref{sup})
 there correspond soft--susy breaking trilinear couplings of the form
\begin{equation}
A_Q h_Q^{ij}Q_iu_j^cH_2 + A_D h_D^{ij}Q_id_j^cH_1
+ A_L h_L^{ij}L_ie_j^cH_1+ B \mu H_1H_2
\end{equation}
where now $A_n = (\bar A_n + 3\bar B_n ) m_{3/2}$, with $n=Q,D,...$
 and $B=(\bar A_{\mu} + 2\bar B_{\mu}) m_{3/2}$.

The  RG--improved effective potential for the neutral particles
can be written  as follows
\begin{eqnarray}
V_0(Q)&=&(m_{H_1}^2+\mu ^2)\mid H_1\mid ^2+
(m_{H_2}^2+\mu ^2)\mid H_2\mid
^2+m_3^2(H_1H_2+h.c.)\nonumber  \\
&+&\frac{g^2+g^{\prime ^2}}8(\mid H_1\mid ^2 - \mid H_2\mid ^2)^2
+\Delta V_1  - \eta(Q)m^4_{3/2}
\end{eqnarray}
where\cite{grz}
\begin{eqnarray}
\Delta V_1=\frac {1}{64\pi^2}Str{\cal M}^4
ln(\frac{{\cal M}^2}{ Q^2} - \frac{3}{2} )
\end{eqnarray}
is the one loop contribution, necessary to stabilize the
minimization  with respect to the higgs vevs
$v_i=<H_i>$, while
\begin{equation}
\label{strace}Str{\cal M}^2(z,\bar z)=\sum_n(-1)^{2s_n}(2s_n+1)
				       m_n^2(z,\bar z)
\end{equation}
is the sum over all particles with spin $s_n$ and mass $m_n$.

The term $\eta (Q)m^4_{3/2}$ comes from the
underlying supergravity theory and can be interpreted
\cite{kpz} as a contribution to the cosmological
constant. Notice that terms of $ {\cal O}(m_{3/2}^2M_{Pl}^2)$
are not included since such contributions would destroy the
hierarchy $m_{3/2}\ll M_{Pl}$.
Specific examples with these desirable features may be
found in four dimensional string models \cite{FKZ}.

The quantum corrections to the above potential depend on
the particle spectrum masses which are computed with the
use of R.G.E.s in terms of the initial values at the unification scale.
A precise knowledge of the R.G. effects in the low
energy theory is important not only from the
phenomenological point of view (measurements of
$sin^2\theta_W, \alpha_{em}$, top mass, e.t.c.) but for
theoretical reasons as well. In particular the Higgs
mass parameters and squark masses of the third
generation play a very important role in the radiative
breaking of the electroweak symmetry on the one hand and
the preservation of the $SU(3)$ confining group and the
electric charge on the other.

{}From phenomenological analyses, two limiting cases have
been of particular interest. The first, is when the two
higgs {\it vev}'s are of the same order $(tan\beta \sim{\cal O}
(1))$ while the difference between the top mass and the
other fermion masses is attributed to a large top Yukawa
coupling relative to the bottom,  $h_t\gg h_b$.

It is interesting that this case can easily be treated
analytically both in the fermion and scalar mass sector
at the one loop level \cite{ilbks,.}.
Since $h_t$ is the only non-negligible Yukawa, all the
others can be ignored and the relevant R.G.E.s decouple,
leading to first order differential equations which can be
solved by standard methods. The same
procedure can be used for the scalars including those of
the third generation.

The case with a large bottom Yukawa coupling
$(tan\beta\gg 1)$ is much more complicated.
In a previous work \cite{..}, however, it was shown that
if one neglects the small differences in the $U(1)_Y$
factors, an analytic solution of the $h_t-h_b$ coupled
differential system can be obtained easily. In
particular, in the absence of a large tau Yukawa
coupling, one can write the $h_t-h_b$  equations
at the 1-loop as follows
\begin{eqnarray}
\frac{d}{dt} h^2_t &=&
\frac{1}{8\pi^2}
 \Big\{6h^2_t + h^2_b - G_Q \Big\} h^2_t \label{eq:1} \\
\frac{d}{dt} h^2_b &=& \frac{1}{8\pi^2}
\Big\{h^2_t + 6h^2_b - G_B \Big\} h^2_b \label{eq:2}
\end{eqnarray}
 with
\begin{equation}
G_Q = \sum^{3}_{i=1} c^i_Q g^2_i\,\, ,\qquad  G_B = \sum^{3}_{i=1}
 c^i_B g^2_i
\label{eq:3}
\end{equation}
where $c^i_Q = \Big\{\frac{16}{3}, 3, \frac{13}{15}\Big\}$ and
$ c^i_B = \Big\{\frac{16}{3}, 3,
\frac{7}{15}\Big\}$ for $SU(3), SU(2)$
and $U(1)$ respectively.

In order to solve Eqs(\ref{eq:1}-\ref{eq:2}), we
define the new parameters $x,y$ through the relations
 $h^2_t = \gamma^2_Q x ,\,\,   h^2_b =
\gamma^2_Q y $, with
\begin{eqnarray}
\gamma^2_Q &=& \exp \Big\{\frac{1}
{8\pi^2} \int_{t}^{t_0}  G_Q({t^{\prime}})
dt^{\prime}\Big\} \nonumber \\
&=&\prod_{j=1}^3\left(1-\frac{b_{j,0}\alpha_{j,0}}{2\pi}
(t-t_0)\right)^{\frac{c_Q^j}{b_j}}    \label{eq:gammaQ}
\end{eqnarray}
 and make the following transformation
\begin{equation}
u=\frac{k_0}{(x-y)^{5/6}}\equiv \frac{k_0}{{\omega}^{5/6}},
\qquad d\,I=\frac{6}{8\pi^2}\gamma_Q^2 d\,t \label{uvar}
\end{equation}
where the parameter $k_0 = 4x_0y_0/(x_0-y_0)^{7/6}$ depends on the
initial conditions $x_0\equiv h_{t,0}^2$ and $y_0\equiv h_{b,0}^2$.
 Then, we can form a differential equation for the new  variable $u$
 which can be cast in the form \cite{..}
 \begin{equation}
\frac{u^{1/5}d\,u}{\sqrt{1+u}}=-\frac{5}{6}k_0^{6/5}d\,I \label{intu}
\end{equation}
 Then, by simple integration it can be shown that
the solution can be given in terms of hypergeometric functions,
 ${}_2F_1(a ,b ,c; z)$, i.e.
\begin{equation}
u^{7/10}{}_2F_1(\frac{1}{2},\frac{-7}{10},\frac{3}{10},
\frac{1}{- u},)
-u_0^{7/10}{}_2F_1(\frac{1}{2},\frac{-7}{10},\frac{3}{10},
\frac{1}{-u_0},)
=\frac{7}{12}k_0^{6/5}I(t)  \label{ansol}
\end{equation}
with $I(t) =\frac{3}
{4\pi^2} \int^t_{t_0} \gamma_Q^2({t^{\prime}})
dt^{\prime}$.
Thus, for a given set $x_0,y_0$ of initial conditions at $M_U$
we can calculate the $u,\omega (t)$ values from equation (\ref{ansol})
and determine the $h_t,h_b$ from
\begin{eqnarray}
h_t^2 &\equiv & \gamma^2_Q x\\  \nonumber
         &=&\gamma^2_Q
\frac{1}{2}\omega (1+\sqrt{1+k_0\omega^{-5/6}})\\
h_b^2&\equiv &  \gamma^2_Q y\\  \nonumber
&=& \gamma^2_Q
\frac{1}{2}\omega (-1+\sqrt{1+k_0\omega^{-5/6}})
\end{eqnarray}
In ref \cite{kpz2} similar solutions where proposed.
A further study of the solutions by these authors showed that all
 $h_{{t,b}_0} >1$ initial values  accumulate on a curve
on the $h_t,h_b$ plane.  In the region $h_t\approx h_b$,
this  curve is slightly deformed when the difference of the
hypercharge  in the two RG--equations is taken into account.
A numerical fit has shown the relation \cite{kpz2}
\beq
( h_t - 0.015)^{12} + ( h_b^2 + 0.045)^{12} =
(\frac{\gamma_Q}{\sqrt{I/2}})^{12}
\eeq
The main deformation occurs around the diagonal, i.e.
when $h_t \sim h_b$. The top mass solutions presented in
the subsequent section are slightly affected by this change. In
any case, this uncertainty on the top mass is in fact overwhelmed
by much bigger uncertainties arising from sparticle loop
corrections on the bottom mass. A more detailed discussion
of the latter will be presented in the next section.
For most applications, (in particular when $u\gg 1$),
it is appropriate to expand the hypergeometric
function and obtain a simplified form of the above solutions.
They can be  expressed in terms of a simplified form of the
function $\omega (t)$,
\beq
\omega (t) = \frac{x_0-y_0}{\{_2F_1^0 + \frac 76
\sqrt{x_0y_0}I(t)\}^{12/7}}
\label{wm}
\eeq
where the value of the hypergeometric function at $u=u_0$ is
$_2F_1^0 \approx 1$ for the limit of interest, i.e. for $h_{t,0}
\sim h_{b,0}$.
Formula (\ref{wm}) is going to be used later in the calculations
for the scalar masses. Note finally (for later use)
the useful relation between $x,y$ variables
\begin{equation}
\Big(\frac {x-y} {x_0-y_0}\Big)^7 = \Big(\frac{xy}
{x_0y_0}\Big)^6 .  \label{eq:xy}
\end{equation}
In particular, in certain supergravity theories the Yukawas
are subject to various constraints which may be combined
with the above formulae and result to interesting
predictions \cite{kpz,bd}.


The analytic $h_t-h_b$ solution neglects small
differences arising from the different hypercharge
assignment of the top and bottom quarks.
This approximation is quite reasonable since other
uncertainties (neglect of Yukawa couplings, one loop
approximation e.t.c) of the same order are also
introduced in the analytic solution.
In any case, it is clear that the vastly different
values of $m_t$ and $m_b$ masses cannot be attributed to
effects due to these small corrections unless a
considerable fine tuning of the parameters occurs.

The usefulness of the analytic expressions for $h_t-h_b$
couplings in the large $tan\beta$ regime, is largely
exhibited in the case of the scalar mass calculations. As
has already been mentioned, the knowledge of the scalar
mass parameters affected by the large Yukawa couplings at
low energies, is very crucial and decisive for the
particular supergravity or string model under
investigation \cite{var}.
To be more concrete, let us assume that only $h_t,h_b$
Yukawas are large. In this case the up and down squark
masses of the third generation as well as the two higgs
mass parameters of the standard model $(m^2_{H_i},i=1,2)$
doublets receive large negative contributions
proportional to integrals involving $h_t^2,h_b^2$.
If $h_b\ll h_t$, and $h_t$ is large enough, then
$m^2_{H_2}$ mass parameter turns to a negative value at
low energies triggering the $SU(2)_L \otimes U(1)_Y$ breaking
radiatively.
On the other hand, $m^2_{H_1}$ mass does not receive
large negative corrections since $h_b$ coupling is
small. This is of course desirable since (as it turns out in
this case), the neutral Higgs potential is stable and the
{\it vev}'s of the Higgses are finite as expected. Besides it
can be proved \cite{.} that for the experimentally accepted top
quark mass and in the largest part of the ($m_{1/2}, m_{3/2}$)
parameter space the squared masses of the top-squarks receiving
negative contributions are not driven negative, protecting in
this  way the color and charge quantum numbers.

The situation is much more complicated in the case of
large bottom coupling too. Then the other Higgs mass
$m^2_{H_1}$ may also be driven negative at low
energies. In order to make the analysis more
transparent, let us write the conditions obtained from
minimization of the Higgs potential, with respect to
$\upsilon_i$'s
\begin{eqnarray}
m_3^2&=&-\frac 12(m_{H_1}^2+m_{H_2}^2+2\mu ^2)\sin 2\beta
\\
m_Z^2&=&2 \frac{m_{H_1}^2-m_{H_2}^2\tan ^2\beta }
{\tan ^2\beta -1}-2 \mu ^2
\end{eqnarray}
In the large $tan\beta $ scenario that we are
interested here $(tan\beta\gg 1)$ we may approximate
the above formulae so that $m^2_Z\simeq -2\,( m^2_{H_2}+\mu ^2)\,$
and $\,- m^2_3\, tan\beta \simeq m^2_{H_1} + m^2_{H_2} +
2 \mu ^2 > 0$.
 Combined together,
these equations lead to $m^2_{H_1} -m^2_{H_2}>
m^2_Z\,$   and
$\mu_1^2\equiv \mu^2+m^2_{H_1} >\frac{m^2_Z}{2}>0\,$ with
 the conclusion that the stability of the potential is
ensured only for a positive value of $\mu_1^2$ mass.
Another compelling reason is also the protection of the
color and charge quantum numbers. The above description
shows clearly the necessity of obtaining analytic
solutions of the scalar mass parameters in the case that
$tan\beta\gg 1$.

A final comment we would like to make concerning the
usefulness of the proposed semi -- analytic procedure is
also based on theoretical motivations. Indeed, as we have already
mentioned, a particularly interesting idea is that the vacuum
energy density could be minimized with respect to the
gravitino mass $m_{3/2}$ in order to fix the mass of the
latter -- and therefore the supersymmetry breaking scale --
dynamically.
The minimization of the vacuum energy with respect to the
$m_{3/2}$ parameter leads to a certain condition for the
renormalised scalar mass parameters and Higgs mass terms
at low energies, which reads \cite{kpz}
\begin{equation}
2V_1(Q) + \eta (Q)m^4_{3/2}=0
\end{equation}
while the coefficient $\eta (Q)$ obeys a renormalisation
group equation of the form
\begin{equation}
\frac{d\, \eta(Q)}{d\, log(Q)} = - \frac{1}{128\pi ^2}
Str{\cal M}^4
\end{equation}
The dynamical determination of the $m_{3/2}$ scale
requires the knowledge of the evolved soft $(mass)^2$--parameters
involved in the above constraint. Thus, analytic knowledge
of their dependence on the boundary
conditions are very useful.
In particular, in the case of non -- universal boundary
conditions of the scalars at $M_U$, the parameter space at
$M_U$ can be represented by a vector $ \vec  \xi =(\xi
_{H_1},\xi _{H_2},\xi _{Q},...)$ where
$\tilde	m^2_i(M_U)=\xi_i\,m^2_{3/2}$. It is not easy to handle
numerically the above system in this case and therefore an
approximate analytic solution of the scalar mass
parameters would be very important.

In what follows we wish to extend our previous results of
the $t-b$ analytic expressions. We give expressions for
various useful quantities including $tan\beta $ and
derive simplified formulae for $h_{t,b}$ Yukawa
couplings. In particular, in the next section we develop the
formalism which is going to be used in section 4 for the derivation
of the scalar mass analytic expressions.

\newpage

\section{Phenomenological analysis of $b-t$ solutions}

Let us define two new parameters $\sigma$ and $\tau$
\begin{eqnarray}
 \tau  &=& exp\Big\{\frac{3}{4\pi^2}
 \int^t_{t_0}  h_t^2\, dt^\prime\Big\} \label{tau} \\
 \sigma  &=& exp \Big\{\frac{3}{4\pi^2}
\int^t_{t_0}  h^2_b dt^\prime\Big\}
\label{sigma}
\end{eqnarray}
The $h_t$-$h_b$ RGE's can be combined to give
\beq
\frac{\dot \tau}{\dot \sigma}
= \frac{x_0}{y_0}\, \left(\frac{ \tau}{ \sigma}\right)^{\frac{11}6}
\eeq
where dots stand for derivatives with respect to $t=ln(Q)$.
Integration of the above gives
\beq
\frac{1}{x_0}\frac{1}{\tau^{5/6}}-\frac{1}{y_0}\frac{1}{\sigma^{5/6}}=
\frac{1}{x_0}-\frac{1}{y_0}\label{int}
\eeq
Rewriting  now  $h_t - h_b$ RGE's in terms of ($\sigma$,$\tau$) and
substituting into (\ref{int}) we obtain
\barr
\tau\sigma &=& \frac{x-y}{x_0-y_0}\nonumber \\
           &=& \left(\frac{u_0}{u}\right)^{\frac 65}
\err
where $u_0\equiv u(x_0 , y_0)$.
On the other hand,
 we can express the ratio $\frac{\tau}{\sigma}$ in terms
of the integral
\barr
\frac{\tau}{\sigma}&=
&exp\Big\{\frac{3}{4\pi^2}\int_{t_0}^t(h_t^2-h_b^2)d\,t\Big\}
\nonumber \\
           &\equiv & exp\Big\{\int_{t_0}^t(x-y)d\,I\Big\}
\err
which, with the help of (\ref{intu},\ref{uvar}), can be expressed as
follows
\beq
\frac{\tau}{\sigma}=\left(\frac{y_0}{x_0}\,
\frac{\sqrt{1+u}+1}{\sqrt{1+u}-1}\right)^{\frac 65}
\eeq
The explicit dependence of $\sigma ,\tau$ parameters on the new
variable $u$
defined in (\ref{uvar}) is not only necessary for the scalar mass
solutions we
are going to discuss in the next section;
 it will also prove  useful in other low energy parameters.
Let us for example consider the evolution of the ratio
of the two higgs vev's ( $tan\beta$), in the case we are examining
here, i.e. when only $h_{t,b}$ couplings are large.
In this case we have the following evolution equation
\beq
\frac{d}{d\,t}tan\beta =
-\frac 14 \frac 3{4\pi^2}(h_t^2-h_b^2) tan\beta
\eeq
Integration of the above gives
\barr
tan\beta &=& tan\beta_0 \left(\frac{\sigma}{\tau}\right)^{\frac 14}
\nonumber \\
    &=& tan\beta_0 \left(\frac {h_{t,0}}{h_{b,0}}\right)^{\frac 35}
            \left(\frac{\sqrt{1+u}-1}{\sqrt{1+u}+1}\right)^{\frac
{3}{10}}
\label{tanb}
\err
Thus $tan\beta$ evolution is also determined solely from the
parameter $u$ and the initial values of the two Yukawas.
{}From (\ref{tanb}) it is obvious that for large $u$ values
the $u$--dependent term in (\ref{tanb}) tends to unity and the
 ratio $R_{\beta}=tan\beta/tan\beta_0$  is given approximately
 by  $(h_{t,0}/h_{b,0})^{3/5}$.

In the most general case, it would be interesting to combine the above
result with the $( m_t , m_b )$  mass relation to eliminate the
parameter $u$.
Notice that to obtain this relation in the case of $tan\beta \gg 1$
 one should take into account corrections to the bottom mass,
arising  from loop--graphs\cite{BL,HRS,copw} containing supersymmetric
 particles.
It is possible however to assume the existence of PQ- or R- type
symmetries
which may prevent the appearance of such large corrections\cite{HRS}.
For the sake of simplicity we will assume that due
 to some kind of symmetry such corrections
are absent. In this case one finds
\beq
m_t = m_b \frac {h_{t,0}}{h_{b,0}}{tan\beta_0}
\left(\frac {tan\beta_0}{tan\beta}\right)^{\frac 23}
\eeq
where $m_b$ is the running bottom mass at the scale $Q=m_t$, related
to the bottom mass at $Q\sim m_b$ with a renormalisation group factor
$\eta_b\approx 1.41$, i.e., $m_b(m_b)=\eta_b m_b(m_t)$.
In figure {\it 1} we show the variation of $r_{\beta}\equiv
(tan\beta/tan\beta_0 )^{2/3}$
 {\it vs} the parameter {\it u}
whose range $(u_0, u_f)$ is determined from the initial
values of the couplings $h_{t,0},h_{b,0}$ and the solution
(\ref{ansol}).
($u_{f}$ corresponds to the value that $u$ receives at the scale $m_t$).
Since we are interested for large $tan\beta$ values, we plot $r_{\beta}$
for several $h_{t,0}/h_{b,0}$ -- ratios close to unity. In all cases we
observe a small change of $r_{\beta}$ which approaches  asymptotically
its
maximal value, $(h_{t,0}/h_{b,0})^{\frac 25}$, as $u\ra \infty$.
 Therefore, in these cases putting all together, one finds the simple
 relation
\barr
m_t\approx m_b(m_t) tan\beta
\err
where  $m_b(m_t)$ is the   $m_b$ mass  at the scale $Q =m_t$.

In Table I, we present selected values of the top mass, in the case
where the top Yukawa coupling is close to its non--perturbative value,
($h_{t,0}=3.0$) and various values of the $t/b$ Yukawa ratio close to unity,
$h_{t,0}/h_{b,0}\ge 1$, i.e. when  $h_{b,0}$ Yukawa coupling is also
large. Both, running and physical top masses are shown while
in the last column we present the corresponding values of
$tan\beta $, assuming a central value for the bottom quark, i.e.
$m_b(m_b) =  n_b \frac{\upsilon} {\sqrt{2}} h_b \cos{\beta} =4.25 GeV$.
Here, as already explained,  $ n_b$ includes the running from the scale
$Q\sim m_t$ down to the scale $m_b$ and is taken to be  $ n_b\approx 1.41$.
We start the running at the unification scale $M_U$, using as inputs
the unification scale itself and the value of the common gauge coupling
$\alpha_U$ at $M_U$ together with the initial values  of the Yukawa
couplings $h_{t,0},h_{b,0}$ as shown in the Table I. Then,
in obtaining the low energy values of $\alpha_{em}$, $a_3$, and
$sin^2\theta_W$, we use the following ranges
\begin{eqnarray}
{\alpha_{em}}^{-1} = 127.9\pm .1 , \,\,   a_3=.12\pm .01 ,\,\,
sin^2\theta_W=.2319\pm  0.0004 \nonumber
\end{eqnarray}
{}For the large $h_{t,0}$ values chosen here, the top mass is in
remarkable agreement with the recent experimental data \cite{cdf}.

Of course the accuracy of the results of the table should not be
overestimated.
A more precise calculation  should involve uncertainties arising
from the strong coupling measurements and
other parameters as the bottom corrections mentioned earlier.
Our aim here is to show  that our solutions result to stable
$m_t,\, m_t^{ph},\, tan\beta$ predictions  against
 $h_{t,0}/h_{b,0}$ variations.


\begin{center}{
\vglue 0.4cm
\begin{tabular}{ccccc}    \hline

$h_{t,0}$ & $ {h_{t,0}}/{h_{b,0}} $ & $m_t/GeV $
 & $ m_t^{ph}/GeV $  & $ tan{\beta} $  \\
 \hline
$3.0$ &$  1.17$  &$176.7$ &$184.7$ &$ 58.22$\\
      &$  1.25$  &$177.0$ &$184.9$ &$ 58.10$\\
      &$  1.30$  &$177.1$ &$185.0$ &$ 58.06$\\
      &$  1.36$  &$177.2$ &$185.2$ &$ 57.90$\\
      &$  1.42$  &$177.3$ &$185.3$ &$ 57.78$\\
\end{tabular}
\vglue 0.2cm
{\bf Table I.} {\it $m_t$ and $\tan{\beta}$ values for
$h_{t,0}=3.0$ and various $h_{t,0}/h_{b,0}$--ratios shown
in the second column.  We fix $m_b  = n_b \frac{\upsilon}
{\sqrt{2}} h_b \cos{\beta} = 4.25 GeV$. Large sparticle loop
corrections on the bottom mass may change the above predictions.}
}\end{center}
Indeed, as has been already noticed, uncertainties from sparticle
loops contributing to $m_b$ are not included.  Therefore, the
analytic solutions  of the scalar masses (presented in the subsequent
section) which are involved in the relevant loops may considerably
simplify the examination of the above corrections. The $\delta m_b$
corrections will have a direct impact on the determination of the angle
$\beta$ and therefore on the top--mass. The relevant loop--graphs\cite{BL}
 lead to the following
result\cite{HRS,copw}
\begin{eqnarray}
\delta(m_b) & = &\mu \frac{\tan\beta}{16 \pi^2}
 (\frac{8g_3^2}{3 } m_{\tilde{g}}
I(m_{\tilde{b},1}^2, m_{\tilde{b},2}^2,
m_{\tilde{g}}^2) +
h_t^2 A_t  I(m_{\tilde{t},1}^2, m_{\tilde{t},2}^2,\mu^2)),
\end{eqnarray}
where the function $I(x,y,z)$ is given by
\begin{equation}
I(x,y,z) = \frac{x y \ln(x/y) + y z \ln(y/z) + x z \ln(z/x) }
{( x - y )( y - z )( x - z )},
\end{equation}
In the above, $m_{\tilde{g}}$ is the gaugino mass, while
$m_{\tilde{b},i},m_{\tilde{t},i}$  are the eigenmasses of
sbottom and stop. A crucial role is played by the relative sign
of the two terms involved  in the above formula.
The gaugino mass is given by its one loop
formula, $m_{\tilde{g}}=\frac{a_3}{a_G}m_{1/2}$. Our analytic
procedure in the next section can also provide  a simple formula
for the trilinear coupling
$A_t=  f_t(A_{t,0},A_{b,0})-I^{\prime}_{\tau}m_{1/2}$,
where $f_t$ is a linear function of $A$'s while $I^{\prime}_{\tau}$
an integral to be discussed in the subsequent section.
{}For reasonable $A_{{(t,b)},0}$ initial values $A_t$ can be
approximated
$A_t \approx -I^{\prime}_{\tau}m_{1/2}$. Thus, there is a
relation between
the trilinear parameter and the gaugino mass, $A_t\approx -0.6
m_{\tilde{g}}$ which can simplify our discussion.
Therefore, there is a partial cancelation of the above two terms
in the formula, but in the   case tan$\beta \approx (50-60)$
as the table shows, the correction is still large.

A precise estimation depends on the details of the scalar mass
spectrum involved in the integral functions
 $I(x,y,z)$ and the specific initial conditions
of the scalars. This is nevertheless  possible for any set of
initial conditions on the scalars with the help of the analytic
solutions
of the next section.  Furthermore the sign of $\mu$ plays also a
decisive role.
It can be easily seen that such corrections can lie in
a wide range, from a  few  up to 40 $\%$.
Evidently, similar graphs to those
considered above, can also result to corrections on the tau mass.
Nevertheless,
in this case it is found\cite{copw} that they are at most $5\%$ and do
not affect essentially the top mass prediction.
It is worth noting here that in particular unified models as in the
case of $SU(5)\times U(1)$\cite{aehn}, the $b-\tau$ Yukawa couplings
 are not necessarily equal at the Unification scale.
Thus, if these corrections come with the right sign,
they may  reconcile large top mass values, with the condition
$h_b > h_{\tau}$. In such cases, our present analytic procedure
where the $h_{\tau}$ coupling is ignored in valid within all
the parameter space.

\newpage

\section{Analytic Solutions for the $3^{rd}$
 Generation Scalar Masses}

We have already pointed out that in unified supersymmetric models
where only $h_t,h_b$ couplings are large, the differential equations
determining the evolution of the soft scalar masses of
the third generation and the two Higgs mass parameters $m_{H_{1,2}}$
are coupled. In the general case with non-zero trilinear
parameters $A_t, A_b$, these equations take the form \cite{var}
\begin{equation}
\label{dmqdt}
\frac{d\tilde m_{Q_L}^2}{dt}=
-\frac 1{8\pi ^2}\sum c_i^QM_i^2g_i^2 +
\frac{h_t^2}{8\pi ^2}{\cal M}_U^2 +
\frac{h_b^2}{8\pi ^2}{\cal M}_D^2 +
			 \frac 16\frac{\alpha _1}{2\pi }S
\end{equation}\nonumber
\begin{equation}
\label{dmudt}
\frac{d\tilde m_U^2}{dt}=
-\frac 1{8\pi ^2}\sum {c_i^UM_i^2g_i^2}+
2\frac{h_t^2}{8\pi ^2}{\cal M}_U^2 -
			  \frac 23\frac{\alpha _1}{2\pi }S
\nonumber
\end{equation}
\begin{equation}
\label{dmhdt}
\frac{dm_{H_2}^2}{dt}=
-\frac 1{8\pi ^2}\sum {c_i^H M_i^2g_i^2}+
3\frac{h_t^2}{8\pi ^2}{\cal M}_U^2 +
			  \frac 12\frac{\alpha _1}{2\pi }S
\end{equation}
\begin{equation}
\label{dmbdt}
\frac{d\tilde m_D^2}{dt}=
-\frac 1{8\pi ^2}\sum {c_i^DM_i^2g_i^2}+
2\frac{h_b^2}{8\pi ^2}{\cal M}_D^2 +
			  \frac 13\frac{\alpha _1}{2\pi }S
\nonumber
\end{equation}
\begin{equation}
\label{dmh1dt}
\frac{dm_{H_1}^2}{dt}=
-\frac 1{8\pi ^2}\sum {c_i^H M_i^2g_i^2}+
3\frac{h_b^2}{8\pi ^2}{\cal M}_D^2 -
			  \frac 12\frac{\alpha _1}{2\pi }S
\nonumber
\end{equation}

where
\begin{eqnarray}
{\cal M}_U^2\equiv \tilde m_{Q_L}^2+
			  \tilde m_U^2+m_{H_2}^2+A_t^2\\
{\cal M}_D^2\equiv \tilde m_{Q_L}^2+
			  \tilde m_D^2+m_{H_1}^2+A_b^2
\end{eqnarray}
and the initial condition for $S$, in the case of MSSM, is given by
\begin{equation}
S_0={(m_{H_2}^2-m_{H_1}^2)}_0+
\sum_{gen}{(\tilde m_Q^2+\tilde m_D^2+
\tilde m^2_E - \tilde m_L^2-2\tilde m_U^2)}_0
\end{equation}
In the case of the universal boundary conditions the $S$ contribution
vanishes identically as required by the absence of gravitational mixed
anomaly\cite{MR}.
In this case $S$ vanishes at any scale as can be deduced
by integrating its corresponding renormalisation group equation,
$S(t)=\frac{\alpha_1(t)}{\alpha_{1,0}}Tr\left[Ym^2\right]$.
It is obvious from the above equations that if we start
with the same or almost equal $h_t,h_b$ couplings,
 the two Higgses are going to evolve in a similar way.
As discussed in the case of Yukawas, again small
differences of $U(1)$ factors in the scalar mass system
cannot be enough to generate a difference of $m^2_{H_1}$,
and $m^2_{H_2}$ running masses at $m_W$. Here, however, in
contrast to the $h_t-h_b$ case, one has two possibilities
of generating $m^2_{H_1} - m^2_{H_2} > m_Z^2$ at $m_W$. The
first is due to the difference of the Yukawa couplings
$h_{t,b}$. In fact the low $tan\beta$ regime is the
limiting case of this possibility. One uses for economy
universal b.c.'s for scalar masses at $M_U$, therefore
$m^0_{H_1} = m^0_{H_2}$ and the difference arising at $m_W$ is
only due to the fact that $h_t\gg h_b$. Here, if $h_b\sim
{\cal O}(h_t)$, universal conditions cannot possibly
lead to $m_{H_1}^2 \not = m^2_{H_2}$, at least not for
the biggest part of the parameter space.
A complete analysis of the parameter space
$(m_{1/2},m_{3/2},\mu ,...)$ at $M_U$ is obviously required to
determine those regions which are compatible with the
radiative symmetry breaking and the phenomenological
expectations. The isolation of the RSB--compatible
regions of $(m_{3/2},m_{1/2})$ will have a further
consequence in low energy phenomenology: it will evidently
 constrain  the region of
the supersymmetric sparticle spectrum masses. Such a
constrained situation may serve as a guide in the
experimental searches for their detection.

Up to know, analytic solutions exist only in the case
where $h_t\gg h_{b,\tau ,...}$\cite{ilbks,.}.
In order to solve the above equations in the case of
 $h_t\sim h_{b}$, we form
the differential equations for two sums of the above
system, namely for ${\cal M}_U^2,{\cal M}_D^2$. Here we will work
only  in the limiting case  where $A_{b,t}\rightarrow 0$.  In fact,
by recalling the same arguments used in the solution of scalar masses
for the case $h_{t,0}\gg h_{b,0}$\cite{.}
 we can conclude that in the IR--limit the contributions
of the trilinear parameters $A_t, A_b$ do not play important role in the
final solutions for the scalar masses as long as $m_{1/2}\le m_{3/2}$.
Therefore, to simplify the subsequent analysis, we drop  $A_U, A_D$
terms (the extension of the solution to the most general case is
straightforward).  It is easily then observed that one can write the
equations for the sums of scalar masses in the following form:
\begin{eqnarray}
 \frac{d{\cal M}_U^2}{dt}  &=&
\frac{1}{8\pi^2}
 \Big\{6{\cal M}_U^2 h^2_t +{\cal M}_D^2h^2_b - G_U^0 m_{1/2}^2\Big\}
\label{eq:28} \\
\frac{d{\cal M}_D^2}{dt}  &=& \frac{1}{8\pi^2}
\Big\{{\cal M}_U^2h^2_t + 6{\cal M}_D^2h^2_b - G_D^0  m_{1/2}^2\Big\}
 \label{eq:29}
\end{eqnarray}
where $G_U^0 = G_Q + G_{H_2} + G_{U^c}$ and $G_D^0 = G_Q + G_{H_1} +
G_{B^c}$.
In this final system, we also observe that the $S$ contribution is no
longer
present in the equations. Indeed, the two sums formed above include the
partners of trilinear terms of the superpotential.
Due to the $U(1)$ invariance of the Yukawa
Lagrangian, the $S$ term is identically zero in the sums.

To simplify the above coupled equations we make the following
transformations
\begin{eqnarray}
\begin{tabular}{ccccc}
${\cal M}_U^2$  &=& $\tau {\cal X}$  ,
 $\tau $ &=& $exp \Big\{\frac{3}{4\pi^2}
 \int^t_{t_0}  h_t^2dt^\prime\Big\} $\label{eq:30} \\
${\cal M}_D^2$ &=& $\sigma {\cal Y}$ ,
 $\sigma  $&=& $exp \Big\{\frac{3}{4\pi^2}
\int^t_{t_0}  h^2_b dt^\prime\Big\}$
\label{eq:30ab}
\end{tabular}
\end{eqnarray}
Then Eqs(\ref{eq:28}-\ref{eq:29}) can be written in the form
\begin{eqnarray}
\tau \frac{d{\cal X}}{dt} &=& \frac{1} {6}  \frac{d\sigma}{dt}{\cal Y}
- \frac{G_U^0}{8\pi^2} m_{1/2}^2 \label{eq:31} \\
\sigma \frac{d{\cal Y}} {dt} &=& \frac{1} {6}  \frac{d\tau}{dt}{\cal X}
- \frac{G_D^0}{8\pi^2} m_{1/2}^2
\label{eq:32}
\end{eqnarray}
Using the formulae of the previous section,
the above equations can be written as follows
\begin{eqnarray}
 \frac{d}{dt}
\left(\begin{array}{c} {\cal X} \\ {\cal Y} \end{array}
 \right) = {\cal H}(t) \left(\begin{array}{c} {\cal X} \\ {\cal Y}
\end{array}
 \right)- \frac {m_{1/2}^2}{8\pi^2} \left(\begin{array}{c}
\frac{G_U^0}{\tau} \\ \frac{G_D^0}{\sigma} \end{array}
\right).
\label{eq:32a}
\end{eqnarray}
with
\begin{eqnarray}
{\cal H}(t)
=\gamma_Q^2(\sigma \tau )^{\frac 7{12}}
\left[\begin{array}{cc}0&y_0(\frac{\sigma}{\tau})^{\frac{17}{12}} \\
x_0(\frac{\tau}{\sigma})^{\frac{17}{12}} &0
\end{array}
 \right] \label{eq:3}
\end{eqnarray}
We may use Green's functions techniques to solve this non--homogeneous
system. For this we need first the solution of the homogeneous part.

If we define  the function $h(u)$ as follows
\beq
h(u)=\left(\frac{x_0}{y_0}\right)^{\frac 65}
\left(\frac{\sqrt{1+u}-1}{\sqrt{1+u}+1}\right)^{\frac{17}{10}}
\eeq
the homogeneous part becomes
\begin{eqnarray}
d\,{\cal X}=-\frac{1}{10}\frac{d\,u}{\sqrt{u^2+u}}h(u) {\cal Y}(u)\\
d\,{\cal Y}=-\frac{1}{10}\frac{d\,u}{\sqrt{u^2+u}}\frac 1{h(u)} {\cal
X}(u)
\end{eqnarray}
There is a rather complicated dependence on $u$ through $h(u)$ ,
however, it is possible to use new parameters which may
simplify the system. For example, if we define
\beq
d\,Q= -\frac 1{10}\frac {d u}{\sqrt{u^2+u}} \label{eqQ}
\eeq
and
\beq
\frac 1{\sqrt{1+u}} = sin 2\phi
\eeq
we can easily find that
\beq
Q = \frac 15 ln(tan \phi) \label{eqphi}
\eeq
Now the homogeneous system becomes
\begin{eqnarray}
\frac{d\,{\cal X}}{d\,Q}=\tilde h(Q) {\cal Y}(Q)\label{dexy1}\\
\frac{d\,{\cal Y}}{d\,Q}=\frac 1{\tilde h(Q)} {\cal X}(Q)\label{dexy2}
\end{eqnarray}
If we seek solutions of the form
\beq
{\cal X}= Exp(\int {\tilde h} A dQ), \,\, {\cal Y}=
Exp(\int \frac B{\tilde h} dQ)
\eeq
then it can be shown that the  unknown functions $A, B$ satisfy the
relation  $B = 1/A$
while we find that $A$ obeys the first
order Differential Equation
\beq
\frac {d\,A}{dQ} = A ( \frac{1}{{\tilde h} A} - {\tilde h}A)\label{eqa}
\eeq
The latter, is a first order Riccati type Equation of the
general form $\dot A + A^2 - f(x) = 0$ whose solutions are known
only for particular forms of the function $f(x)$.
In our case, $x=\int {\tilde h}(Q) d\, Q$ and
$f(x) = 1/{\tilde h}(Q)^2$, which is a rather complicated
function of the new parameter $x$.
Therefore exact analytic solutions
are possible only in particular limiting cases.
We can study the above equation by geometric methods and obtain useful
information about its solutions. Such a mathematical analysis, however,
is beyond  the purposes of this paper.\footnote{see for example
\cite{soc}}
We will present  soon, two limiting cases  which are
of particular interest from the physics point of view.
Before we proceed to the presentation of the interesting limiting cases
let us exhibit some interesting properties of the equation (\ref{eqa}).
The above equation  remains invariant when $A$ is
shifted by a function $g(Q)$, i.e.  $A \rightarrow A + g$,
which obeys the equation
\begin{eqnarray}
g(Q) &=& \frac{g_0 Exp(-2\int A {\tilde h} d\,Q)d\,Q
}{1 + g_0 \int {{\tilde h} Exp(-2\int A {\tilde h} d\,Q)d\,Q}}
\nonumber \\
&=&\frac{g_0}{1+g_0 \int {\tilde h}
{\cal X}^{-2}d{Q^\prime }} \frac 1{{\cal X}^2}
\end{eqnarray}
Furthermore, the equation $A(u)^2-h(u)^2=0$ defines a curve which
separates
the areas where the derivative of $A(u)$ changes sign. Therefore above
(below)
that curve, $A(u)$ is an increasing (decreasing) function of the
parameter $u$.
This is depicted in figure {\it 2}.
Let us also discuss the possibility of searching for a simplified
solution
of the system (\ref{dexy1},\ref{dexy2}).
 We observe from figure {\it 2} that for approximately
equal couplings, the function $h(u)$ is approximately constant
 within the biggest range of the parameter $u$.
  In figure {\it 3}
we compare the numerical solution for the homogeneous part of the
differential
system with the one obtained by considering $h(u)\approx const.$,
assuming
non--universal boundary conditions.
We can conclude that the approximate analytic solution is in  good
agreement
with the numerical one for $h_t/h_b \le 1.1$. In what follows, we will
consider approximated formulae in the interesting case of approximately
equal $top-bottom$ Yukawas. In terms of the variable $Q$ the complete
system
is written
\begin{eqnarray}
\frac{d}{dQ}
\left(\begin{array}{c} {\cal X} \\ {\cal Y} \end{array}
 \right) =
 \left[\begin{array}{cc}0&\tilde h(Q) \\
\frac1{\tilde h(Q)} &0
\end{array}
 \right]
 \left(\begin{array}{c} {\cal X} \\ {\cal Y}
\end{array}
 \right)- \left(\begin{array}{c}
{G_U} \\ {G_D} \end{array}
\right).
\label{eq:32Q}
\end{eqnarray}
where
where $G_{U,D}(Q)$ are properly defined through (\ref{eq:32a}).

With the use of the diagonalizing matrix,
\begin{eqnarray}
{\cal U}(\tilde h)
= \frac 1{\sqrt{2}\tilde h}
\left[\begin{array}{cc}\tilde h& - \tilde h \\
1 &1
\end{array}
 \right] \label{eq:3}
\end{eqnarray}
 we can write the Differential Equations as follows
\begin{eqnarray}
\frac{d}{dQ}
\left(\begin{array}{c} \tilde {\cal X} \\ \tilde {\cal Y} \end{array}
 \right) = \left\{
 \left[\begin{array}{cr}1&0 \\
 0 & - 1
\end{array}
 \right] -\frac {17}{2\sqrt u}
\left[\begin{array}{cc}1&1 \\
 1 &1
\end{array}
 \right] \right\}
 \left(\begin{array}{c} \tilde {\cal X} \\ \tilde {\cal Y}
\end{array}
 \right)- {\cal U}^{-1}\left(\begin{array}{c}
{G_U} \\ {G_D} \end{array}
\right).
\label{eq:32Q}
\end{eqnarray}
(where the new vector $\Omega^T =(\tilde {\cal X}, \tilde {\cal Y}) $
is related to the old one $\Omega^T_0 =( {\cal X}, {\cal Y}) $
with the matrix ${\cal U}$, $\Omega = {\cal U}^{-1}\Omega_0$).
For the case under investigation, for example, this form is
particularly useful. In the limit $h_{t,0}\approx h_{b,0}$,
the parameter $u\gg 1$, thus the second term  in the right hand
side may be considered as a  first order perturbation.
There is another interesting limiting case which can be also easily
treated, i.e., when $\sqrt{u}\ll 17$ in the range of integration.
In that case (which corresponds to $h_t\gg h_b$), we may define
the new variable $P=ln(h)$ and the transformation
\begin{eqnarray}
{\cal V}
= \frac 1{\sqrt{2}}
\left[\begin{array}{cr}1&1 \\
1 &-1
\end{array}
 \right] \label{eq:v}
\end{eqnarray}
the differential system can be written as
\begin{eqnarray}
\frac{d}{dP}
\left(\begin{array}{c} \hat {\cal X} \\ \hat {\cal Y} \end{array}
 \right) =\left\{
 17 \left[\begin{array}{cc}1&0 \\
 0 & 0
\end{array}
 \right] +{\sqrt u}
\left[\begin{array}{cc}0&1 \\
 1 &0
\end{array}
 \right] \right\}
 \left(\begin{array}{c} \hat {\cal X} \\ \hat {\cal Y}
\end{array}
 \right)- {\cal V}^{-1}{\cal U}^{-1}\left(\begin{array}{c}
{G_U} \\ {G_D} \end{array}
\right).
\label{eq:32Q}
\end{eqnarray}
This particular case is treated more easily if one ignores $h_b$
coupling.
Particular expressions are found in the literature\cite{ilbks,.}.
Thus, in the present work we restrict ourselves in the approximate
bottom--top
equality. It can be easily
concluded in that case that $u\gg 1$ always and the function $h(u)$
 is  approximately constant, $h(u)\approx
(\frac {x_0}{y_0})^{6/5}$. In this case
 Eq(\ref{eq:32Q}) can be integrated, leading to the following result
\begin{eqnarray}
\left(\begin{array}{c} {\cal X} \\ {\cal Y} \end{array}
\right) =Exp\left[{\cal H}_0(Q -
Q_0)\right]\left\{\left(\begin{array}{c}
{X_0} \\ {Y_0} \end{array}
\right)
+\int_{Q_0}^{Q} dQ^{\prime}Exp\left[-{\cal H}_0(Q^{\prime} - Q_0)\right]
  \left(\begin{array}{c}
{G_U} \\ {G_D} \end{array}
\right)\right\}.
\label{eq:32L}
\end{eqnarray}
where, due to the property ${\cal H}^2\equiv
\left(\begin{array}{cc} 1&0 \\ 0&1 \end{array}
 \right)$,
the exponents can be expanded as follows
\begin{eqnarray}
Exp\left[{\cal H}_0(Q - Q_0)\right]=
\left(\begin{array}{cc}
ch(Q - Q_0)&h_0 sh(Q - Q_0)\\
h_0^{(-1)}sh(Q - Q_0)&ch(Q - Q_0) \end{array}
\right)
\end{eqnarray}
In the large $tan\beta$ regime and taking into account the $Q(u)$
dependence
of Eq(\ref{eqQ}), for the present purposes the above solutions can be
approximated as follows
\begin{eqnarray}
\frac{{\cal M}_U^2}{m_0^2}
&\approx
&\frac{\tau}{2} \left\{(\xi_{U}+h_0\xi_{D})\rho +
(\xi_{U}-h_0\xi_{D})\frac{1}{\rho}\right\}+\xi_{1/2}<I_{\tau}>
\\
\frac{{\cal M}_D^2}{m_0^2}
&\approx
&\frac{\sigma}{2 h_0} \left\{(\xi_{U}+h_0\xi_{D})\rho -
(\xi_{U}-h_0\xi_{D})\frac{1}{\rho}\right\} + \xi_{1/2}<I_{\sigma}>
\end{eqnarray}
with $\rho = (\frac {tan\phi}{tan\phi_0})^{(1/5)}$.
 In the last two expressions
we have used the convenient parametrisation ${\cal M}_{(U,D)_0}^2=
\xi_{(U,D)} m_0^2\,$ and $\, m_{1/2}^2 = \xi_{1/2}m_0^2\,$, while
$\xi_U\equiv \xi_{H_2} +\xi_Q + \xi_{t^c}$ and
$\xi_D\equiv \xi_{H_1} +\xi_Q + \xi_{b^c}$.
Finally the integrals $<I_{\tau ,\sigma}>$ are given by
\beq
<I_{\tau}>=\frac{\tau}2 \int_t^{t_0}\frac
{G_U({t^\prime})}{\tau{({t^\prime})}}
\left\{(1+\nu{(t^\prime)})\frac 1{\rho (t^\prime)}+(1-\nu{(t^\prime)})
\rho (t^\prime)\right\}
d{t^\prime} \label{Itau}
\eeq
and similarly for $<I_{\sigma}>$ with the replacements $\tau\ra \sigma $
and $G_U \ra G_D$, $\nu \ra \frac 1{\nu}\,$ while $\nu $ is given by
\beq
\nu  = \left(\frac {\sqrt{1+u}-1}{\sqrt{1+u}+1}\right)^{\frac 12}
\eeq
The integrals $<I_{\tau ,\sigma}>$ look complicated, but in fact they
can easily be performed using  definitions (\ref{uvar},\ref{wm}) to
express  the functions $\rho , \nu , \tau , \sigma$ in terms of the
integration variable $t^\prime $. As a matter of fact, for the case
of interest ($h_b\sim h_t$), $u \gg 1$ the integrals can be simplified
further, as $\nu \ra 1$. Then, instead of (\ref{Itau}), we simply write
\beq
<I_{\tau}>\approx \tau \int_t^{t_0}\frac {G_U({t^\prime})}
{\tau{({t^\prime})}\rho{({t^\prime})}}dt^{\prime} , \;\;
<I_{\sigma}>\approx \sigma \int_t^{t_0}\frac {G_D({t^\prime})}
{\sigma{({t^\prime})}\rho{({t^\prime})}}dt^{\prime}
\eeq
Once the solutions of the sums are obtained, we may substitute  back
into the equations for the individual masses. A simple way to solve
these equations is achieved if we formally integrate the sums to obtain
\begin{eqnarray}
{\cal M}_U^2 - {\cal M}_{U,0}^2 - C_U(t)m_{1/2}^2 = - 6 J_U -
J_D\label{al1}\\
{\cal M}_D^2 - {\cal M}_{D,0}^2 - C_D(t)m_{1/2}^2 =  - J_U - 6
J_D\label{al2}
\end{eqnarray}
with $J_I(t) = \int h_I^2 {\cal M}_I^2 dt$, $I=U,D$. Now, the unknown
integrals
$J_I(t)$ can be expressed in terms of the already calculated sums,
their
initial conditions and known gauge functions, from the simple algebraic
system
(\ref{al1},\ref{al2}). Then, the higgs mass parameters for example can
be
given from
\begin{eqnarray}
m_{H_1}^2 = (\xi_{H_1} + C_H(t)\xi_{1/2}) m_0^2 - 3 J_D(t) +
I_S\label{hig2}
\\
m_{H_2}^2 = (\xi_{H_2} + C_H(t)\xi_{1/2}) m_0^2 - 3 J_U(t) - I_S
\end{eqnarray}
with $I_S$ representing the integral of the $S$--contribution in the
case of non -- universality. Similar expressions hold also for the
other scalar masses of the third generation.

In order to examine the radiative  breaking of the electroweak
symmetry, we are finally interested for the mass parameters
$\mu_i^2 = m_{H_i}^2+\mu^2$.
The initial value of $\mu$ is also unknown, but this arbitrariness
can be eliminated by using the minimization conditions and determine
its value in terms of known quantities.
For the large $\beta$ we can write
\beq
\mu^2 \approx - m_{H_2}^2+\frac{m_{H_1}^2}{tan^2\beta}-\frac 12 m_Z^2
\label{eqmu}
\eeq
Notice that with the use of the relevant renormalization group equation
for $\mu$ and (\ref{eqmu}), we may determine the initial condition for
the $\mu$ parameter which may be useful in several cases.

Let us conclude this section with some remarks concerning
the advantage of the above semi--analytic procedure with respect
to the standard numerical integration.
First we have shown that the useful quantities of the minimal
susy model (Yukawa solutions, scalar masses, tan$\beta$,
$\mu$--term e.t.c.)  can be expressed in terms only of
 known gauge functions and the parameter $u$.
The value of the latter can be obtained  easily at any scale in
terms of  the top and bottom couplings at the
unification scale through the hypergeometric function defined in
(\ref{ansol}). In fact, in the limiting case we discuss  in this
work,  the parameter $u$ can be expressed by a simple, approximately
linear relation in terms of its initial value, thus a lot of
information about the low energy parameters can be easily obtained
by a straightforward calculation of the involved  parameters.
Furthermore, the scalar spectrum depends on two more new quantities,
i.e. the integrals $I_{\tau , \sigma}$, which can also be easily
computed or fitted by simple mathematical functions for any given
$h_{(t,b)_0}$ set.

A final comment we wish to make here concerns the possible effects of
the $A_{t,b}$ parameters in our solutions. We have already noted that,
the $A_t-A_b$  renormalisation group equations
can be solved independently in terms of the gauge and Yukawa couplings.
In fact, the $A_{t,b}$ coupled differential system  is exactly the same
with that of the scalar sums  (\ref{eq:29}) with the proper replacement
of the coefficients $G_{U,D}$. After replacing their solutions
$A_{t,b}= f_{t,b}(A_{t,0},A_{b,0}) - I^{\prime}_{\tau ,\sigma}m_{1/2}$,
into the system of the scalar masses, one finds that the $\delta_{A_i}$
corrections are of the form (see also \cite{cwltop,fll})
\beq
\delta_{A_i} = \delta_{i_1} A_0^2 + \delta_{i_2}A_0m_{1/2}+
\delta_{i_3}m_{1/2}^2
\eeq
The coefficients $\delta_{i_{(1,2,3)}}$ involve the two more integrals
similar to $I_{\tau , \sigma}$ defined above. Remarkably, in the case
of Yukawa couplings close to their infrared value, the coefficients
multiplying the $A_0$ terms are found to be small $\delta_{i_{1,2}}\ll 1$.
This result has also been confirmed elsewhere \cite{cwltop}, where
all terms  involving the initial condition $A_0$
have been shown to be multiplied by a factor
$1-(\frac {h_t}{h_t^{fxd}})^2$, with $h_t^{fxd}$ the fixed point
value of the top Yukawa coupling. For reasonable $A_0,m_{1/2}$
values, (i.e., not much larger than the gravitino mass),
and staying close to the fixed point $h_t$--value, the above
corrections do not  modify substantially the present analysis.

\newpage

\section{Results and Conclusions }

We are ready now to present our numerical examples using the obtained
scalar mass formulae.
Although the formulae of the previous section look rather complicated,
in fact for most of the cases of interest, they can be simplified
considerably. As a matter of fact, the only ``new'' quantities
that enter the various scalar masses are $J_{U,D}$ which mainly
depend on the two integrals $I_{\tau}, I_{\sigma}$ and the initial
boundary conditions $\xi_i$. For example,  in the case
of $h_t\sim h_b$ couplings close to their non--perturbative regime,
the $J_{U,D}$ quantities can be easily simplified as follows
\begin{eqnarray}
J_U\approx \frac 17 \left\{3.0+\epsilon_{\xi}
 +(C_U(t)-I_{\tau})\xi_{1/2}\right\}m_0^2\label{ju}\\
J_D\approx \frac 17 \left\{3.0+\bar\epsilon_{\xi}
 +(C_D(t)-I_{\sigma})\xi_{1/2}\right\}m_0^2\label{jd}
\end{eqnarray}
where $\epsilon_{\xi},\bar\epsilon_{\xi}$ represent small corrections
of the order ${\cal O}(\sim 3\%)$ which can be easily extracted
from the previous formulae. Furthermore the $C_{U,D}=\sum_nC_n^{U,D}$
 gauge functions are easily calculated from the formulae
\begin{equation}
C_n(t)=\sum^3_{i=1}
\frac{c_i^n}{2b_i\alpha^2_{i_G}}
\left(\alpha_i^2(t_1)-\alpha_i^2(t)\right)
,\quad\quad n=1,2,3
\end{equation}
 Thus $C_U\approx  7.1+6.7+0.5= 14.3$ and similarly for $C_D$.
Finally, for the region of interest the two integrals
$I_{{\tau},{\sigma}}$
are easily calculated. For $h_{t_0}\approx 3.0$ for example we find
$ I_{\tau}\approx 5.7$, while as $h_{b_0} < h_{t_0}$, we obtain
$I_{\sigma } > I_{\tau}$.
In the case of equal bottom--tau  couplings the two expressions
(\ref{ju},\ref{jd}) are almost identical (up to U(1) differences
in $C_{U,D}$ gauge quantities and the
$\epsilon_{\xi},\bar\epsilon_{\xi}$
parameters which might  differ due to non--universal initial boundary
conditions). In the limiting case of equal b.c's at the GUT scale
these quantities are equal and our expressions reduce to those of
similar investigations of references \cite{cwltop,kpz2}.

Let us now turn to the numerical investigation of the scalar masses.
Using the  the minimization conditions for the potential,
 it  has been shown  in section 2  that in order
to achieve Radiative Symmetry Breaking at low energies,  the
constraint  $\delta m_{H_{1,2}}^2\equiv m_{H_1}^2-m_{H_2}^2 \ge m_Z^2$
should be satisfied.
After some simple algebra we arrive at
\begin{eqnarray}
\frac{\delta m_{H_{1,2}}^2}{m_0^2}& =
 & \frac 15 \left\{ 2 (\xi_{H_1}-\xi_{H_2})
     + 3 (\xi_{u^c}-\xi_{d^c})\right\} +2 I_S \nonumber \\
 &-&{\frac 3{10}}\left[(\tau -\frac {\sigma}{h_0})\rho (\xi_U+h_0\xi_D)
 +(\tau +\frac {\sigma}{h_0})\frac {1}{\rho}
 (\xi_U-h_0\xi_D)\right]\nonumber \\
&+&\frac{3}{5}\left[ (C_{u^c}-C_{d^c}) +
 I_{{\sigma}{\tau}}\right]\xi_{1/2}
\label{Dm12}
\end{eqnarray}
with the $I_{{\sigma}{\tau}}$ integral being the difference
$I_{\sigma}- I_{\tau}$.  In the above formulae,
all functions depending on the parameter $u$ are finally expressed
in terms of the initial value $u_0$, which depends only on the ratio
of Yukawas, $h_{t,0}/h_{b,0}$ at $M_{GUT}$.
Therefore, the constraint $m_{H_1}^2-m_{H_2}^2> m_Z^2$
has been reduced down to a simple algebraic inequality of the form
\beq
\alpha \xi_{H_1} + \beta \xi_{H_2} +\gamma \xi_{1/2} +\delta > \xi_Z
\label{dm12}
\eeq
with $\xi_Z = (m_Z/m_0)^2$ while $\alpha ,\beta , \gamma$ and $\delta$
can be
easily read from (\ref{Dm12}). (The other constraints can also be easily
converted to simple equations of this latter form).
Notice that in the limit $h_b \ra h_t$ the integrals $I_{\tau ,\sigma}$
are approximately equal while $C_{u^c},C_{d^c}$ differ only in the
$U(1)$
factors. Therefore, in the case of non -- universality, in general we
get
 $\gamma < \alpha , \beta $ from (\ref{dm12}), and the main dependence
is on $\xi_{H{1,2}}$ parameters.
For $h_t\ge h_b$ at $M_U$ we get $I_{\sigma\tau }\ge 0$,
therefore the coefficient
in front of $\xi_{1/2}$  is always positive. Furthermore, for universal
boundary conditions $\xi_{H_1}=\xi_{H_2}, \xi_{u^c}=\xi_{d^c}$ the first
line in the RHS of equation (\ref{Dm12}) becomes zero.
In this case a lower bound is put on value of $\xi_{1/2}$ to satisfy
(\ref{dm12}). In any case, this condition can be naturally satisfied
for non -- universal boundary conditions and in particular if
 $\xi_{H_1} > \xi_{H_2}, \xi_{u^c}\ge \xi_{d^c}$.
 Thus, once a particular point $(\xi_{H_1} ,\xi_{H_2} ,\xi_{1/2})$
is chosen, this condition may  be seen as a lower bound
on the scale $m_{3/2}$, i.e.
$m_{3/2}\ge m_Z/(\alpha \xi_{H_1}+
\beta\xi_{H_2}+\gamma\xi_{1/2}+\delta)^{1/2}$.
In fig.{\it 4}, this constraint is represented by a two dimensional
surface  as a function of the $ \xi_{H_1}$
and $\xi_{1/2}$ parameters choosing two specific values for $\xi_{H_2}$.
Evidently, for small $\xi_{1/2}$ the constraint is satisfied only for
$\xi_{H_1} > \xi_{H_2}$.

We may further examine  the Higgs mass term $m_{H_2}^2$ which should
be negative in order to break the $SU(2)$ symmetry.
If we substitute $J_U$ from the previous equations in (\ref{hig2}),
(ignoring for simplicity small corrections of the order ${\cal
O}(2-3\%)$), we arrive at the result:
\begin{eqnarray}
\frac{m_{H_2}^2}{m_0^2} &\approx &
 \frac 1{35} \left\{ (17\xi_{H_2}+3\xi_{H_1})-3
( 5\xi_Q +6 \xi_{u^c}-\xi_{d^c})\right\}
\nonumber \\
&-& \frac 37 \{ C_Q + C_{u^c} -\frac 43 C_H - I_{\tau}
\}\xi_{1/2}\label{H2}
\end{eqnarray}
After substituting the relevant functions by their numerical values,
one can easily check under what conditions the required inequality
($m_{H_2}^2 < 0$) is satisfied.
For example assuming universal conditions for all scalars except
$m_{H_1}$, a lower bound is put in the parameter
$\xi_{1/2}$ only if $\xi_{H_1}\ge 4$
i.e., when $m_{H_1}\ge 2 m_{3/2}$.

Similarly, for the $\tilde t^c$ squark running mass we get
\begin{eqnarray}
\frac{\tilde m_{t^c}^2}{m_0^2}
&\approx &
\frac 1{35} \left\{ (23\xi_{u^c} + 2\xi_{d^c})-2
( 5\xi_Q +6 \xi_{H_2}-\xi_{H_1})\right\}
\nonumber \\
&-& \frac 27 \{ C_Q + C_H -\frac 52 C_{u^c} - I_{\tau} \}\xi_{1/2}
\end{eqnarray}

Now, if we denote with $\tilde m_{t^c}^{exp.}$ the lower experimental
bound on the stop--mass, it is rather simple to investigate the
constraint
 $\tilde m_{t^c}^2 > \left(\tilde m_{t^c}^2\right)^{exp.}$
from the above formula. In fact, this can be also converted to
a bound on the parameter $m_{3/2}$ as in the case of Eq.(\ref{dm12}).
In most of the parameter space however, this bound on $m_0^2$
is lower than the corresponding obtained from $\delta m_{H_{12}}^2$
condition. In fact, the $\left(\tilde m_{t^c}^2\right)^{exp.}$ -- bound
excludes only regions with $\xi_{1/2}\ll 1$.

Nevertheless, interesting  bounds
 -- in addition to $\xi_{H_1}>\xi_{H_2}$ -- arise mainly from the
$\tilde m_{b^c}$ -- mass, given by
\begin{eqnarray}
\frac{\tilde m_{b^c}^2}{m_0^2}
&\approx &
\frac 1{35} \left\{ (23\xi_{d^c} + 2\xi_{u^c})-2
( 5\xi_Q +6 \xi_{H_1}-\xi_{H_2})\right\}
\nonumber \\
&-& \frac 27 \{ C_Q + C_H -\frac 52 C_{d^c} - I_{\sigma} \}\xi_{1/2}
\end{eqnarray}
For many interesting regions of the $\xi_i$ parameter space the
bounds obtained from this latter mass are non --trivial. Assume
for example a moderate case where $\xi_Q = \xi_{d^c} = \xi_{H_2}
=\frac 12$. (This would correspond to the relation
$\tilde m_i \approx 0.7 m_{3/2}$ for the relevant mass parameters).
Then the above formula implies the following relation between
$\xi_{1/2},\xi_{H_1}$
\beq
\xi_{1/2}\ge \frac 2{25}\xi_{H_1} +\frac 1{30} \left[7 \left(
\frac{\tilde m_{b^c}^{exp}}{m_0}\right)^2- \frac {17}{10}\right]
\eeq
where $\tilde m_{b^c}^{exp}$ is the experimental
 lower bound of the sbottom mass.
Now, if we imply a sensible value for $m_{H_1}$ (say $m_{H_1} \approx
2 m_{3/2}$) so as to satisfy naturally the condition(\ref{Dm12}), we
should
have  at least $\xi_{1/2}\ge (0.3-0.4)$, no matter how small the ratio
${\tilde m_{b^c}^{exp}}/{m_0}$ is. This in turn implies the constraint
$m_{1/2}\ge (0.55-0.65) m_{3/2}$. Of course this holds for the
particular
set of $\xi_i$'s chosen, but a detailed investigation can be easily done
for
the whole parameter space.

We should finally point out that the above results do not include
contributions from a large $h_{\tau}$--Yukawa coupling which in various
models is also predicted to be of the same order with $h_{t,b}$.
A complete account of these effects in the described analytic
procedure up to now was not possible. However, corrections of
this Yukawa coupling are not expected to have a significant impact
on the large part of the parameter space. Nevertheless,
it will be a rather interesting question if the above analytic
procedure can also be extended to include  corrections of $h_{\tau}$,
in the case where $h_t\sim h_b \sim h_{\tau}$.


In conclusion, in this paper we have presented analytic expressions
for the Higgs mass parameters $m_{H_{1,2}}^2$ and squark masses
of the third generation, when both
top and bottom Yukawa couplings are large, as predicted by many
unified theories. In the derivation of these expressions we
 have used exact analytic solutions of the  top--bottom
($h_t, h_b$) running Yukawa couplings  ( expressed in terms of
 hypergeometric functions) which are involved in the evolution
equations of the scalar masses. We have shown that, in the limit
$h_t, h_b \gg h_{\tau},...$
the coupled differential system of Higgses and scalars  quarks
is reduced down to a simple first order Riccati Differential equation.
Specific numerical examples have been  presented in the
interesting case of large $tan\beta \approx {\cal O}(60)$,
using  values for the top coupling close to its non--perturbative
regime. For these ranges of the parameters $h_t,tan\beta$, and
a bottom mass $m_b\approx 4.25 GeV$, we predict a running top mass
$m_t\approx (176-177) GeV$ corresponding to a physical mass
$m_t^{ph}\approx (184-185) GeV$.
These predictions can change substantially, if corrections from
sparticle exchange loops (being proportional to $tan\beta$)
are taken into account.
They depend crucially on the value of the $\mu$ parameter,
while they are mainly controlled by a linear combination of the
 $g_3^2m_{\tilde g}$ and $h_t^2A_t$ quantities,
where  $m_{\tilde g}, A_t$ are the gaugino mass
and the trilinear coupling respectively. It is possible of course
to assume proper boundary conditions to suppress those corrections down
to a $~(5-6) \;\%$ of the bottom mass,
 but in principle they can be as large as $40\%$
is specific regions of the parameter space. A final answer to this
question can be given only when the underlying theory is known  to
pick up a particular point of the parameter space. For the time being
we can reverse the procedure and constrain the arbitrary parameters
using the experimentally determined range of the top mass.

We have further presented simple expressions for the Higgses
and scalars and discussed the constraints put by the radiative symmetry
breaking mechanism, as well as, by charge and color protection
on the initial conditions of the scalars.
Thus the condition that the one Higgs $(mass)^2$ turns negative at low
energies can be easily satisfied for natural values of the scalar
masses as can be concluded from (\ref{H2}).
Furthermore,  the second Higgs is protected from large negative
corrections at low mass scales, either if we impose
$m_{H_1}^2 > m_{H_2}^2$ at the unification point, or if
$m_{1/2} > m_{3/2}$.
It is evident from (\ref{Dm12}) that the specific boundary conditions on
the squark masses $\tilde m_{t^c},\tilde m_{b^c}$,
will play a role in the exact
determination of the ranges of the above mass parameters.
Finally  $\tilde m_Q,\tilde m_{t^c},\tilde m_{b^c}$ squark masses
also impose additional constraints on the $(m_{3/2},m_{1/2})$
mass parameters.

\newpage



\vfill
\newpage

{\bf FIGURE CAPTIONS}

\vspace*{3cm}

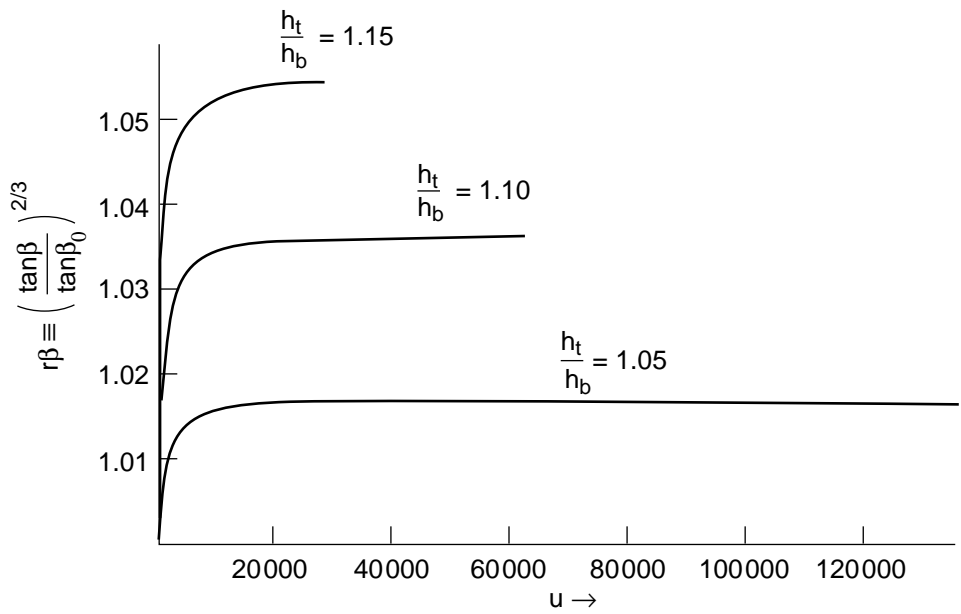
{\it Figure 1.} Plots of the ratio
$r_{\beta}=(tan\beta/tan\beta_0)^{2/3}$ {\it vs}
the  parameter {\it u} for selected ratios of the Yukawa
couplings $h_{t,0}/h_{b,0}$.

\vspace*{1cm}

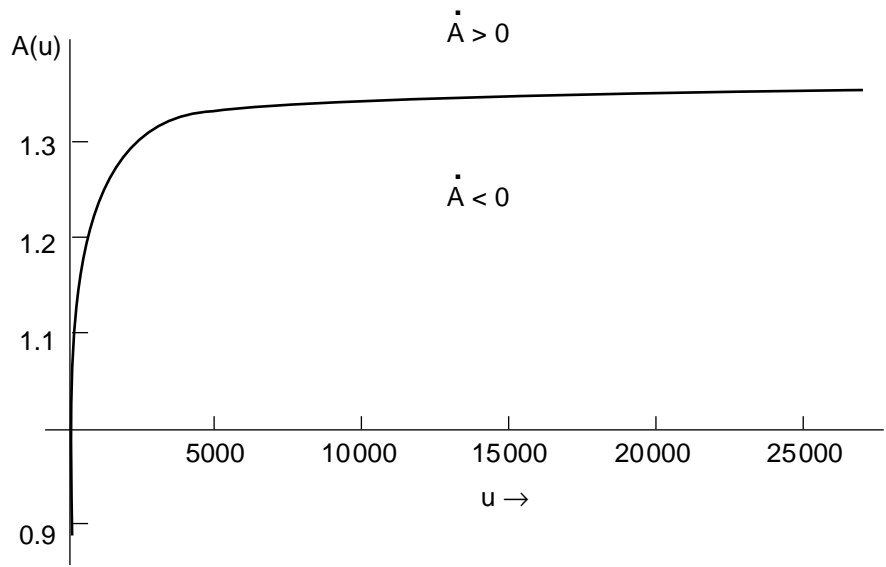
{\it Figure 2.} The curve $A(u) = h(u)$ separating the two regions
with $\dot A > 0$ and  $\dot A < 0$.

\vspace*{1cm}

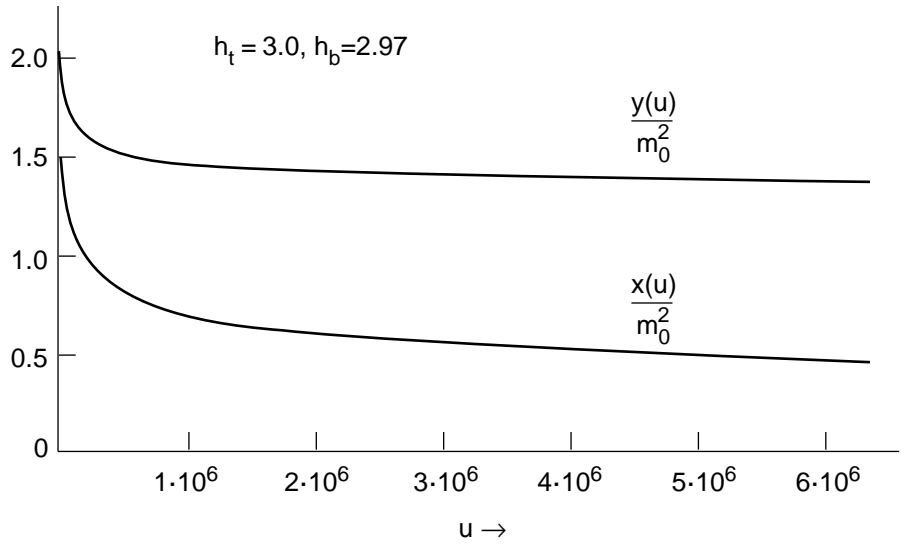
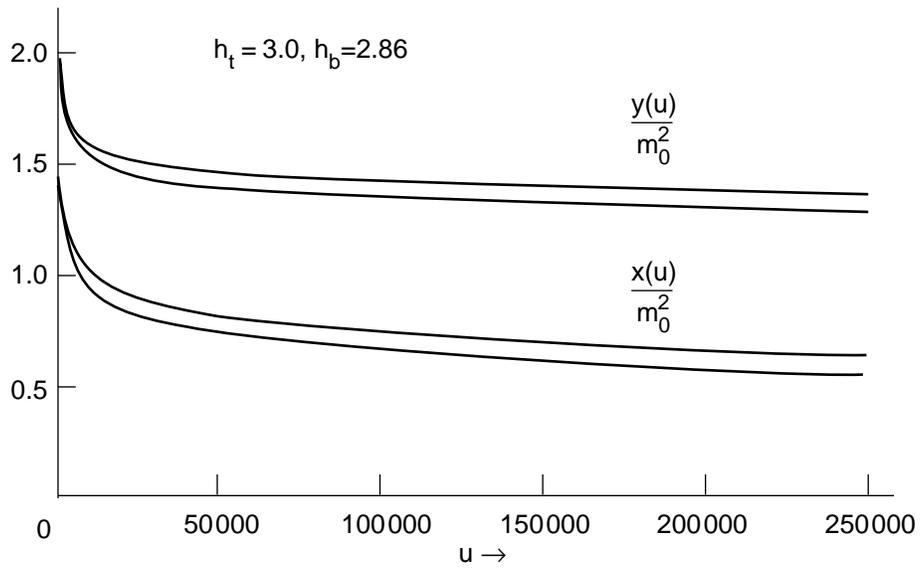
{\it Figure 3.} Comparison of analytic approximate (upper curves)
and numerical solutions (lower curves) of the ratios
${\cal X}(u)/m_0^2,
{\cal Y}(u)/m_0^2$, for two cases of Yukawa couplings. (For ${\cal X},
{\cal Y}$ definitions see section 4).

\vspace*{1cm}

{\it Figure 4.} Surfaces representing the lower $m_0^2$ bound, in
the parameter space $\xi_{1/2}=m_{1/2}^2/m_0^2$, $\xi_{H_1}=
m_{H_1}^2/m_0^2$ for $\xi_{H_2}=0.64$ (lower) and $\xi_{H_2}=1$
(upper case).
If  $\xi_{H_2}\ge \xi_{H_1}$, surface points exist only for
$\xi_{1/2}\gg 1$.
\end{document}